\documentclass[journal=nalefd,layout=twocolumn]{achemso}

\usepackage[utf8]{inputenc}
\usepackage[T1]{fontenc}
\usepackage{color}
\usepackage{amsmath}
\usepackage{amsfonts}
\usepackage{amssymb}
\usepackage{bm}
\usepackage[colorlinks=false, pdfborder={0 0 0}]{hyperref}
\usepackage{graphicx}
\usepackage[justification=justified]{caption}
\usepackage{upgreek}

\newcommand{\myparallel}{{\mkern3mu\vphantom{\perp}\vrule depth 0pt\mkern2mu\vrule depth 0pt\mkern3mu}}

\author{Jan Hajer}
\affiliation[Universität Würzburg]{Institute for Topological Insulators and Physikalisches Institut (EP3), Universität Würzburg, Am Hubland, DE-97074 Würzburg, Germany}
\email{jhajer@physik.uni-wuerzburg.de}

\author{Maximilian Kessel}
\affiliation[Universität Würzburg]{Institute for Topological Insulators and Physikalisches Institut (EP3), Universität Würzburg, Am Hubland, DE-97074 Würzburg, Germany}
\altaffiliation{Present and permanent address: Center for Quantum Spintronics, Department of Physics, Norwegian University of Science and Technology, NO-7491 Trondheim, Norway}

\author{Christoph Brüne}
\affiliation[Universität Würzburg]{Institute for Topological Insulators and Physikalisches Institut (EP3), Universität Würzburg, Am Hubland, DE-97074 Würzburg, Germany}
\altaffiliation{Present and permanent address: Center for Quantum Spintronics, Department of Physics, Norwegian University of Science and Technology, NO-7491 Trondheim, Norway}

\author{Martin P. Stehno}
\affiliation[Universität Würzburg]{Institute for Topological Insulators and Physikalisches Institut (EP3), Universität Würzburg, Am Hubland, DE-97074 Würzburg, Germany}

\author{Hartmut Buhmann}
\affiliation[Universität Würzburg]{Institute for Topological Insulators and Physikalisches Institut (EP3), Universität Würzburg, Am Hubland, DE-97074 Würzburg, Germany}

\author{Laurens W. Molenkamp}
\affiliation[Universität Würzburg]{Institute for Topological Insulators and Physikalisches Institut (EP3), Universität Würzburg, Am Hubland, DE-97074 Würzburg, Germany}

\title{Proximity Induced Superconductivity in CdTe-HgTe Core-Shell Nanowires}


\begin{document}


\begin{abstract}
In this letter we report on proximity superconductivity induced in CdTe-HgTe core-shell nanowires, a quasi-one-dimensional heterostructure of the topological insulator HgTe.
We demonstrate a Josephson supercurrent in our nanowires contacted with superconducting Al leads.
The observation of a sizable $I_c R_n$ product, a positive excess current and multiple Andreev reflections up to fourth order further indicate a high interface quality of the junctions.
\end{abstract}

\noindent\rule{\linewidth}{0.5pt}

The prospect of topological quantum computation has spurred a large-scale research effort in identifying material systems that host topological quantum states.
Emerging from an interplay between band structure topology and superconductivity, topological quantum states can be artificially engineered by coupling a conventional \emph{s}-wave superconductor to a semiconductor with strong spin-orbit coupling subjected to a magnetic field\cite{Lutchyn2010, Oreg2010, Mourik2012, Deng2012, Das2012, Churchill2013, Ren2018, Fornieri2018} or to a material with topological band dispersion.\cite{Fu2008, Fu2009}
Recently, we have presented experimental evidence for the fractional Josephson effect in Josephson devices with weak links made of the topological insulator (TI) material HgTe,\cite{Wiedenmann2016, Bocquillon2017, Deacon2017} indicating the presence of non-localized Majorana modes in these structures.
In an alternative approach it has been suggested to create Majorana wires by gate-defining one-dimensional wires in HgTe quantum wells.\cite{Reuther2013}
In this Letter, we report the first observation of proximity superconductivity in CdTe-HgTe core-shell nanowires,\cite{Kessel2017} a material system in development as a possible alternative to the conventional semiconductor nanowire platform.\par


\begin{figure}[ht!]
\centering
\includegraphics[width=0.40\textwidth]{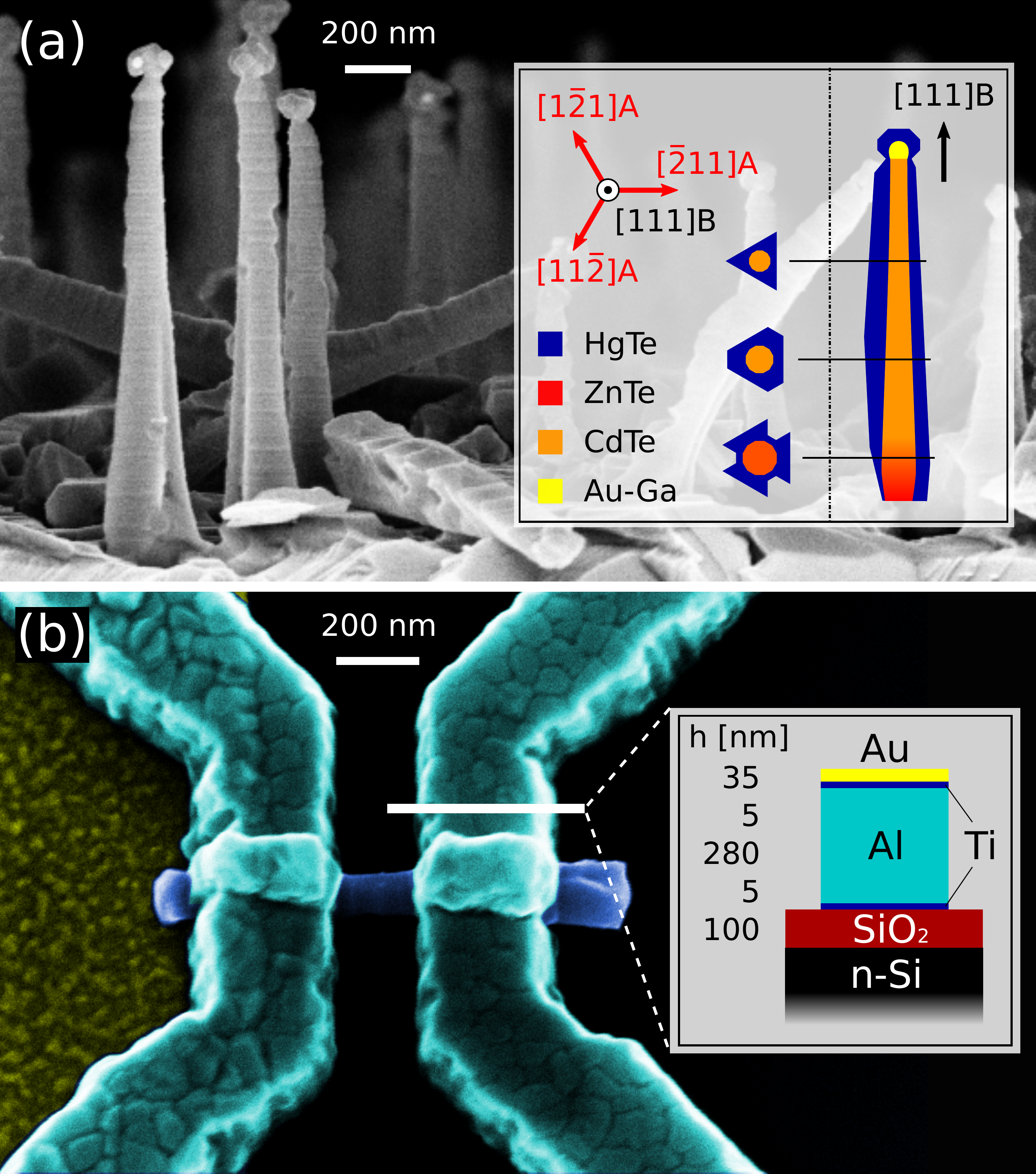}
\caption{Figure (a) shows an scanning electron microscope (SEM) image of CdTe-HgTe core-shell nanowires on (111)B GaAs with schematical cross sections in the inset. An exemplary device is shown as a false-coloured SEM image in (b) with a nanowire (blue) aligned at the DEP electrode on the left and contacted by Al-based leads (turquoise). The inset depicts the contact material stack on the substrate.}
\label{fig_SEM}
\end{figure}

In order to fabricate our core-shell heterostructures, we grow CdTe nanowires oriented along the [111]B direction by vapor-liquid-solid molecular beam epitaxy on (111)B GaAs substrates.
Droplets of Au-Ga are used to seed the one-dimensional growth.
The droplet size defines the nanowire diameter, which is typically 60--80\,nm.
Subsequently, all side-facets of the CdTe wires are overgrown with 30--40\,nm of HgTe.
Further details regarding the growth process are discussed in {Ref.~\cite{Kessel2017}}.
The investigated heterostructures have a high crystal quality and show residual strain due to the lattice mismatch between HgTe and CdTe.\cite{KesselStrain}
Strain lowers the symmetry of the zincblende unit cell and opens a bandgap in HgTe.\cite{Bruene2011}
At low temperature, we expect the CdTe cores to be insulating.
Hence, charge transport takes place in the HgTe shell.\par

In order to contact individual nanowires, we prepare a nanowire suspension in isopropanol by ultrasonic cavitation to transfer them onto a conductive Si substrate capped with 100\,nm of thermal oxide.
We localize and align the wires at a Au electrode using AC dielectrophoresis.\cite{Pohl1951}
We then define the contact pattern by electron-beam lithography.
To achieve good interface quality, the contact area is cleaned from organic residues and oxide by Ar ion milling followed by \emph{in situ} metallization.
A 5\,nm-thick layer of Ti is evaporated as an adhesion layer for the subsequently deposited 280\,nm-thick Al superconductor.
The electrodes are capped with 5\,nm Ti and 35\,nm Au.
A schematic of the layer stack and a scanning electron micrograph are shown in Fig.~\ref{fig_SEM}(b).\par

We studied eleven nanowires in total.
The transport measurements on samples W1--W10 were carried out in a dilution refrigerator at a temperature below 50\,mK.
Sample W11 was tested with specialized low-noise electronics to reduce current-noise-rounding of the current-voltage characteristic.
Further, a microwave-filtered dilution refrigerator setup with a base temperature of 50\,mK was used to eliminate the possibility of spurious resonant features in the dynamic resistance.
DC and standard lock-in measurements were performed in four-terminal geometry.\par~\par

The critical temperature of the superconducting electrodes is found to be $T_c = 1.0\,\text{K}$.
Using the BCS relation $\Updelta=1.764\,k_B T_c$,\cite{Bardeen1957} we estimate an energy gap of $2\Updelta = 0.30\,\text{meV}$.
Due to the geometry of the leads, we observe different critical magnetic fields, $B_{c,\,\myparallel} = 10\,\text{mT}$ and $B_{c,\,\perp} = 21\,\text{mT}$, parallel and perpendicular to the long axis of the nanowire, respectively.\par


\begin{figure}[ht!]
\centering
\includegraphics[width=0.40\textwidth]{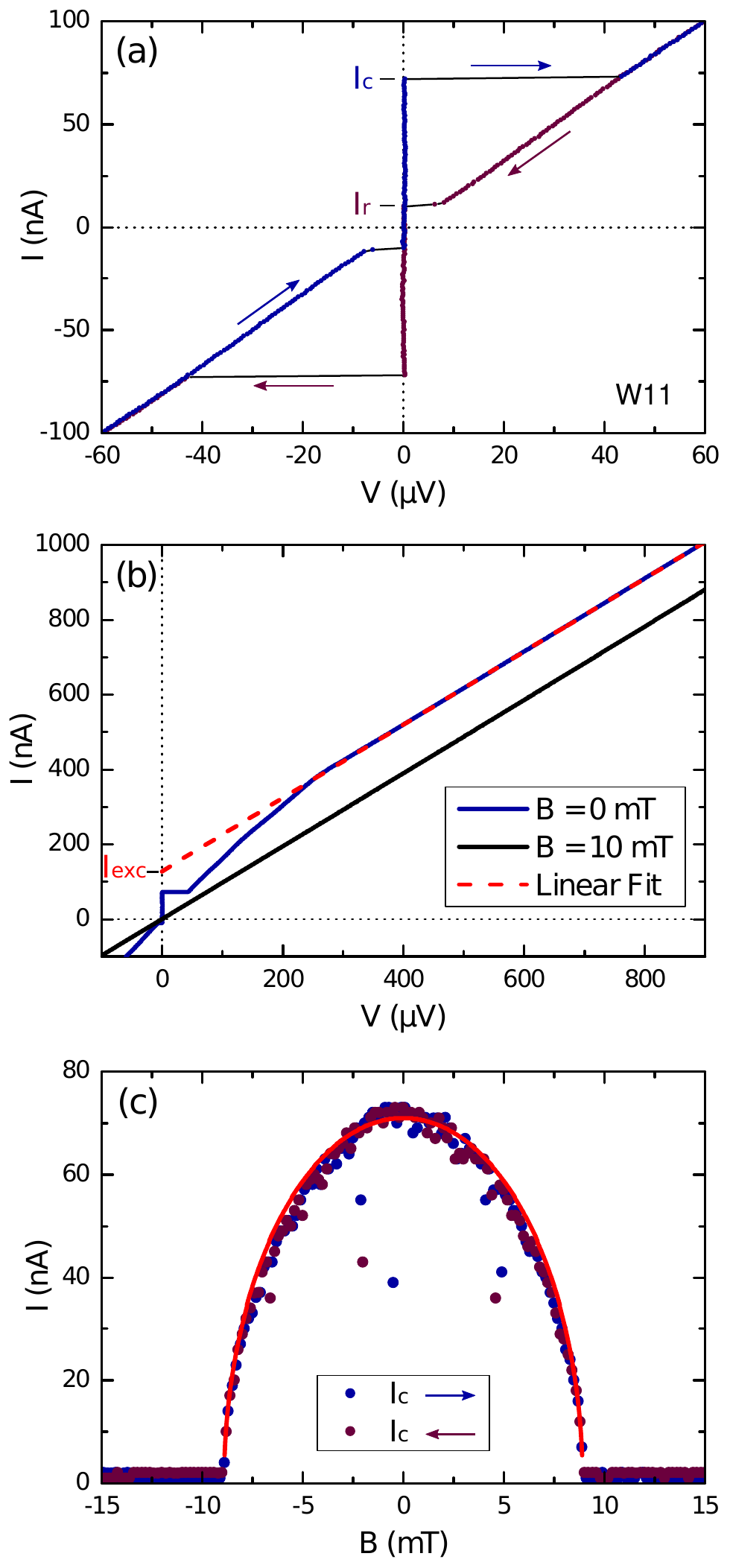}
\caption{The low temperature $I(V)$ characteristic of junction W11 is shown in (a). Arrows and colors indicate different sweep directions. The blue trace in (b) shows the zero-magnetic field $I(V)$ curve for a large extended bias range, with an extrapolation for large bias voltages drawn in red to extract the excess current. The black line shows the ohmic characteristic, where superconductivity is suppressed by an external magnetic field of 10~mT. The magnetic field dependence of the critical current is shown in (c), fitted with equation (\ref{eq1}) (red line).}
\label{fig_IV}
\end{figure}

Figure \ref{fig_IV}(a) shows the current-voltage characteristic of a representative wire junction (W11).
We observe a critical current $I_c = 73\,\text{nA}$ and hysteretic behavior with a retrapping current $I_r = 11\,\text{nA}$.
The normal state resistance of the device, $R_n=1.02\,\text{k}\Upomega$, is extracted from the slope of the current-voltage characteristic for bias voltages $|V|\gg2\Updelta/e$.
This value is identical to the resistance when superconductivity in the contacts is suppressed by a magnetic field. 
The strength of the Josephson coupling is characterized by the product of critical current and normal state resistance; $I_c R_n=75\,\upmu\text{V}$ for sample W11.
It is comparable in magnitude to the experimentally determined gap voltage $\Updelta/e=150\,\upmu\text{V}$ of the electrodes, which indicates a large induced gap in the topological material.\footnote{For comparison, see Table I in Ref.~\cite{Galletti2014}.}
Junction parameters for W1-W10 are reported in the Supplementary Materials.

Extrapolating the current-voltage characteristic at high bias ($V \gg 2\Updelta/e$) to $V=0$ [red dashed line in Fig.~\ref{fig_IV}(a)] reveals a substantial excess current,\cite{Likharev1979} $I_{exc}=128\,\text{nA}$. We attribute the enhanced conductance to quasiparticle transport by Andreev reflections that occur at the transparent interfaces between the superconductors and the TI nanowire.\cite{Klapwijk1982}
In the absence of an adequate theoretical description for the TI nanowire system, we interpret the excess current in the framework of the Octavio, Tinkham, Blonder, and Klapwijk (OTBK) theory\cite{Octavio1983} for symmetric superconductor--normal~conductor--superconductor (SNS) junctions.
Using the expression by Niebler, \emph{et~al.},\cite{Niebler2009} we calculate a BTK barrier parameter\cite{Blonder1982} $Z=0.65$ for W11.
This corresponds to a transparency $T={1}/{1+Z^2}=0.70$, which confirms the high quality of our \emph{ex situ} prepared superconductor--nanowire interfaces.\par


\begin{figure*}[ht!]
\centering
\includegraphics[width=0.75\textwidth]{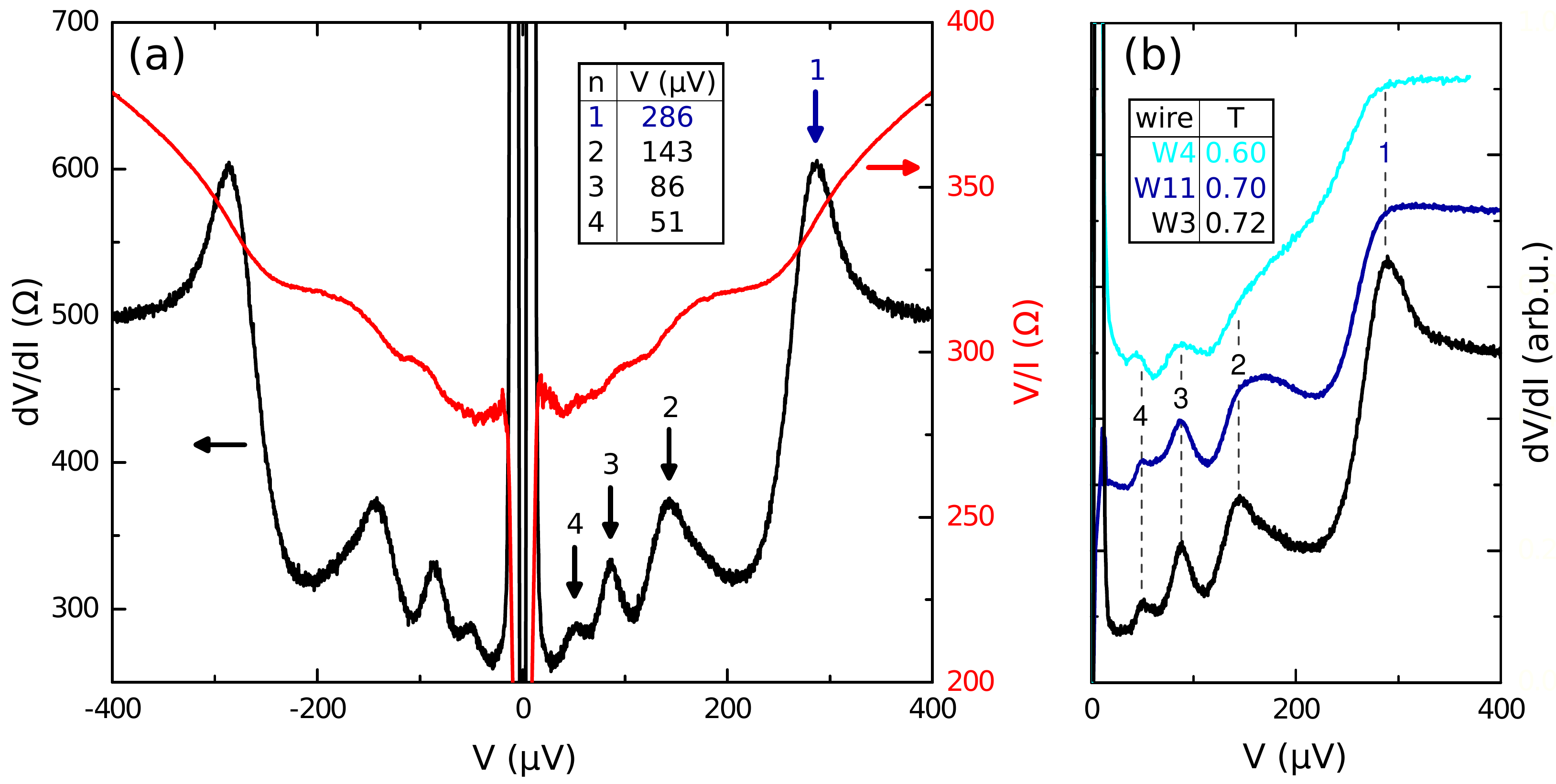}
\caption{Bias dependent resistance $V/I$ (red, right axis) and dynamic resistance $\text{d}V/\text{d}I$ (black, left axis) are plotted in (a). Pronounced features are indicated by numbers and the corresponding bias voltages are listed in the figure. Figure (b) compares the peaks in dynamic resistance for three more wires with different junction transparencies $T$.}
\label{fig_MAR}
\end{figure*}

\begin{figure}[ht!]
\centering
\includegraphics[width=0.5\textwidth]{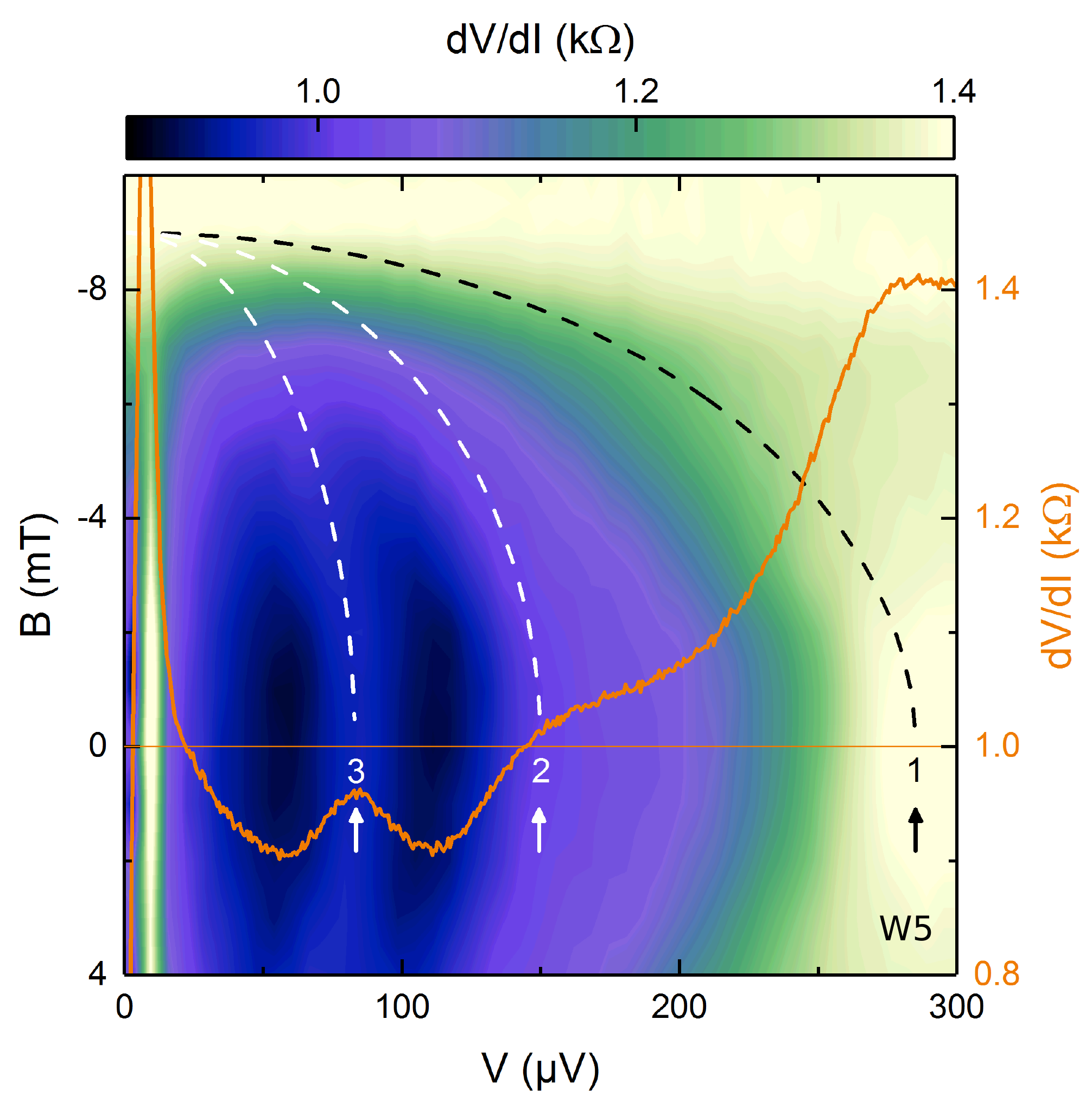}
\caption{The color coded map shows the magnetic field dependence of features in the dynamic resistance $\text{d}V/\text{d}I$ of W5, fitted with equation (\ref{eq1}) (dashed lines). The linescan (orange, right axis) indicates the corresponding peak positions 1--3 for $B=0$.}
\label{fig_Bdep}
\end{figure}

Figure~\ref{fig_MAR}(a) shows the bias voltage dependence of resistance $V/I$ (red) and dynamic resistance $\text{d}V/\text{d}I$ (black). 
We observe a sequence of broad features, which are marked by arrows in the plot.
These features are most pronounced for sample W2 and, with the exception of two samples, all devices display them at the same bias voltages as marked in Fig.~\ref{fig_MAR}(b).
Such subgap structure arises from higher-order quasiparticle transport which involves multiple Andreev reflections (MAR)\cite{Octavio1983} at the interfaces. 
Quasiparticles that transverse the weak link with an applied voltage $V$ gain the energy $e\cdot V$.
In a symmetric junction between superconductors with an energy gap 2$\Updelta$, this leads to a junction conductivity which modulates with $2\Updelta/ne$, where $n=1,2,...$ is the order of the MAR process.  
While the voltage spacing deviates from a standard MAR sequence for $n=3,4$, it does scale with the size of the energy gap $\Updelta(B)$ when superconductivity is suppressed by an applied magnetic field,\cite{Douglass1961}
\begin{equation}
\frac{\Updelta(B)}{\Updelta(0)}=\sqrt{1-\left(\frac{B}{B_c}\right)^2}\ ,
\label{eq1}
\end{equation}
as indicated by the dashed lines in Fig.~\ref{fig_Bdep}, where the magnetic field dependence of features in the dynamic resistance is shown in a color coded map.
Similarly, the critical current scales with $\Updelta(B)$ in the electrodes, see Fig.~\ref{fig_IV}(c).
Strong deviations in the MAR characteristics have been discussed in the context of quantum point contacts with strong coupling to superconducting electrodes.\cite{Dolcini2008}
Disregarding the charging energy, it was predicted that the presence of quantized levels and strong spin-orbit interaction leads to a broadening and shift in the position of MAR features.
We expect strong spin-orbit coupling in our nanowires.
However, the effect of band structure details and channel transmission on Andreev transport in TI nanowires has not been studied theoretically.
Unconventional MAR sequences are reported for asymmetric SNS devices with electrodes of different gap magnitudes\cite{Kuhlmann1994, Guenel2014} and in hybrid devices with proximity induced superconductivity for which additional MAR features arise from the quasiparticle density of states in the proximity region above the effective gap energy.\cite{Galletti2017,Kjaergaard2017}
It should be pointed out that based on the calculations in {Ref.~\cite{Cuevas2006}}, the magnitudes of critical current and excess current $e I_{c} R_n/\Updelta\approx e I_{exc} R_n/\Updelta=0.5$ for W11 are compatible with diffusive SNS junctions in the intermediate regime $L\approx 4\text{--}5\,\xi$, where $L=300$\,nm is the distance between the superconducting electrodes and $\xi$ the superconducting coherence length in the material.
In this case, however, a regular MAR sequence is expected.\par~\par


In conclusion, we successfully demonstrate proximity induced superconductivity in our HgTe nanowire shells contacted with superconducting Al leads.
Apart from observing supercurrent, a sizable $I_c R_n$ product, positive excess current, and a clear signature of a MAR-like subgap structure evidence the high quality of the junctions.
The unusual shape of the subgap dynamic resistance cannot be understood in terms of the standard theory of SNS Josephson devices, but suggests nontrivial interplay with the electronic properties of the HgTe nanowire shell.
This work constitutes an important step towards the realization of topological quantum states in quasi-one-dimensional nanowires of mercury telluride.
Efforts toward adjusting the Fermi energy in the wires by electrostatic gating and testing of electrode materials and geometries with larger critical fields are currently underway.
This will allow us to test for topological superconductivity in the few mode regime and look for a predicted topological transition with applied magnetic flux along the wire axis of a TI nanowire.\cite{Egger2010}


\begin{acknowledgement}
We gratefully acknowledge the financial support of the ERC Advanced Grant (project 4-TOP); the DFG (SFB 1170, SPP1666, Leibniz-Preis); and the Bayerisches Staatsministerium für Wissenschaft und Kunst (ENB IDK TOIS, ITI).
\end{acknowledgement}


\begin{suppinfo}
Overview of device and transport properties of the investigated nanowire junctions
\end{suppinfo}


\providecommand{\latin}[1]{#1}
\makeatletter
\providecommand{\doi}
  {\begingroup\let\do\@makeother\dospecials
  \catcode`\{=1 \catcode`\}=2 \doi@aux}
\providecommand{\doi@aux}[1]{\endgroup\texttt{#1}}
\makeatother
\providecommand*\mcitethebibliography{\thebibliography}
\csname @ifundefined\endcsname{endmcitethebibliography}
  {\let\endmcitethebibliography\endthebibliography}{}

\end{document}


\begin{table}[ht]
    \renewcommand{\arraystretch}{1.5}
    \small
    \centering
    \begin{tabular}{l|ccccccccccc}
        \textbf{Nanowire} & \textbf{W1} & \textbf{W2} & \textbf{W3} & \textbf{W4} & \textbf{W5} & \textbf{W6} & \textbf{W7} & \textbf{W8} & \textbf{W9} & \textbf{W10} & \textbf{W11}\\
        \hline
        $\bm{L_{JJ}}~[\text{nm}]$ & 270 & 270 & 270 & 165 & 195 & 195 & 240 & 195 & 165 & 240 & 270 \\
        $\bm{R_n}~[\text{k}\Omega]$ & 0.99 & 0.51 & 0.57 & 0.77 & 1.39 & 1.62 & 1.34 & 1.80 & 0.88 & 1.99 & 1.02 \\
        $\bm{I_c}~[\text{nA}]$ & 14 & 47 & 37 & 18 & 7 & 8 & 7 & 8 & 18 & 4 & 73 \\
        $\bm{I_r}~[\text{nA}]$ & 14 & 47 & 37 & 18 & 7 & 8 & 7 & 8 & 18 & 4 & 11 \\
        $\bm{I_c\cdot R_n}~[\upmu\text{V}]$ & 14 & 24 & 21 & 14 & 9 & 13 & 9 & 15 & 16 & 8 & 75 \\
        $\bm{I_{exc}}~[\upmu\text{A}]$ & 0.11 & 0.27 & 0.24 & 0.11 & 0.07 & 0.06 & 0.05 & 0.06 & 0.11 & 0.03 & 0.13 \\
    \end{tabular}
    \captionof{table}{Device and transport properties of the investigated Josephson junctions: junction length $L_{JJ}$, normal state resistance $R_n$, critical current $I_c$, retrapping current $I_r$, characteristic voltage $I_c R_n$, and excess current $I_{exc}$. The measurements on wires W1--W10 were performed in a current-noise-limited setup and thus do not exhibit hysteresis in the current-voltage characteristic, i.e.\ $I_c = I_r$.}
    \renewcommand{\arraystretch}{1}
\label{tab_Overview}
\end{table}